# Enhance students' learning outcomes by redesigning individual learning activities into group activities for introductory level Physics courses

Kalani Hettiarachchilage[1] and Neel Haldolaarachchige[2]
[1]Department of Physics and Astronomy, College of Staten Island, New York 10314, USA
[2]Department of Physical Science, Bergen Community College, Paramus, New Jersey 07652, USA

*Abstract*—The evolution of science education is a dynamic process driven by advances in pedagogy, technology, and especially, our understanding of how students learn. Educators are exploring innovative teaching and learning methodologies such as active learning, incorporated technology, interdisciplinary approach, flipped classrooms, personalized teaching, and many more. The goal of all these evolving methodologies is to empower students with not only a strong foundation in scientific knowledge but also with the skills and mindset required to thrive in the future world. By adopting these innovative approaches, educators can help students become effective problem solvers, critical thinkers, and life-learning citizens. Our focus is redesigning individual student learning activities into group learning activities that will benefit various ways of education by building each other's support, connection, and communication. Unloading individual homework and loading in-class group work during a synchronous classroom setting will benefit all levels of learnings effectively. Assigning most of the classwork to be completed during class time has potential benefits since students are more likely to be engaged and focused individually and jointly by sharing and communicating subject matter effectively. We are discussing how to facilitate the group work effectively by creating group assignments for personalized classrooms, setting specific rules and class ethics to follow for the best learning practice, imposing a list of roles to each group member, introducing a list of ground rules related to justice, equity, and diversity inclusion (JDEI) within the group. The students' performances, progress and effectiveness are analyzed and compared. The quality of the work is evaluated.

## I. Introduction

The College of Staten Island (CSI) is a four-year senior college of The City University of New York. CSI is offering Doctoral programs, Advanced Certificate programs, and Master's programs, as well as Bachelor's and Associate's degrees to a diverse student population. Two semester academic patterns with separate summer and winter sessions are followed by the college with the schedule of days, evenings, and weekend classes. As of fall 2022, the student ethnicity/demographics at the college were 44.5% White, 24.8% Hispanic, 13.0% African American, 12.4% Asian, 2.5% Nonresident alien, 2.5% two or more races, 0.1% American Indian or Alaskan Native and 0.1% Native Hawaiian or Pacific Islander [1].

As a primarily undergraduate institution, the college offers associate and bachelor's degree programs. Most of associate and baccalaureate degrees offer in Science and technology, Liberal art and science, School of health science, School of education and School of business divisions are required to complete 100 level Physics classes offered in Physics and Astronomy department. Most students in these classes are struggling a lot and they require more effective innovative teaching methods, encouragement, appreciation, more supportive classroom environment, and strategies to proceed. On the other hand, these are the classes where learners can learn and develop many life-learning skills and strategies reflected in their future. Class designing and restructuring of these key classes should be carefully handled by reflecting course learning and student learning outcomes. We selected Physics I (PHY 116), Physics II (PHY 156) and Introduction to Physics (PHY 114) classes for our investigation for this paper. Every class is filled with around 100 students per semester. These classes are mostly algebra-based gateway physics classes that are offered for students in different disciplines as required to complete for their major degrees. All these classes have both laboratory and lecture components. After covid, we are offering these classes in blended learning approach where the lecture portion is conducted as online mix modality as one large section with all students. The laboratory sessions are held in-person with smaller groups of 18 students. The goal of this paper is to enhance students' learning outcomes of these entire classes by unloading all individual homework assignments to in class group activities by partitioning class time more effectively to learn better.

## II. Methodologies

### A. Class Structure

In 2022, we restructure PHY116, PHY156 and PHY114, 4 credits classes into three components to provide learners with the best and equal education and learning experience [2, 3, 4]. The first component is an asynchronous lecture assigned for one hour class period associated with a simple task to complete. This aims to provide the key content knowledge of the crucial concepts that build the required knowledge for successful continuation of the rest of the class structure effectively. The second component is the problem-solving session which is assigned to a two-hour class period to use learned concepts in e-lectures toward problem-solving and real-world applications. This was studied under mandatory participation in discussions and choice-based participation with required submissions [2]. The third one is an in-person laboratory where we bring students in person for 2 hours per week and connect physics concepts to laboratory applications and demonstrations to provide them with hands-on experiences.

In the lecture portion, students are evaluated with various graded and ungraded assignment categories. There are 42

submissions for the semester including 12 homework assignments. Four-unit quizzes are assigned with synchronous modality with face framed camera on [2, 5]. All class assignments are free response questions and are supposed to be submitted as handwritten work by following personalized class requirements and ethics set up by the instructor to earn full credits. Zero-cost materials, effective feedback, and one-to-one support are provided throughout the semester [2-13]. All classes are well organized and built in learning management system (LMS) for all learners to navigate class materials easily in unit wise modules [14, 15]. The students' anonymous surveys show that 98% of students were able to navigate the page easily.

*B. Class concerns*

All classes are full of registration with around 100 students and students' retention of each class is well preserved (above 95%). Around 95% of students who volunteered for anonymous surveys were happy about the class structure, class materials, LMS usage, what they learned, earned, continuous support, and feedback. Demographic information of the classes is presented in Figure 1 for Spring 23. This information stays mostly the same, closely matched in Fall 23 data.

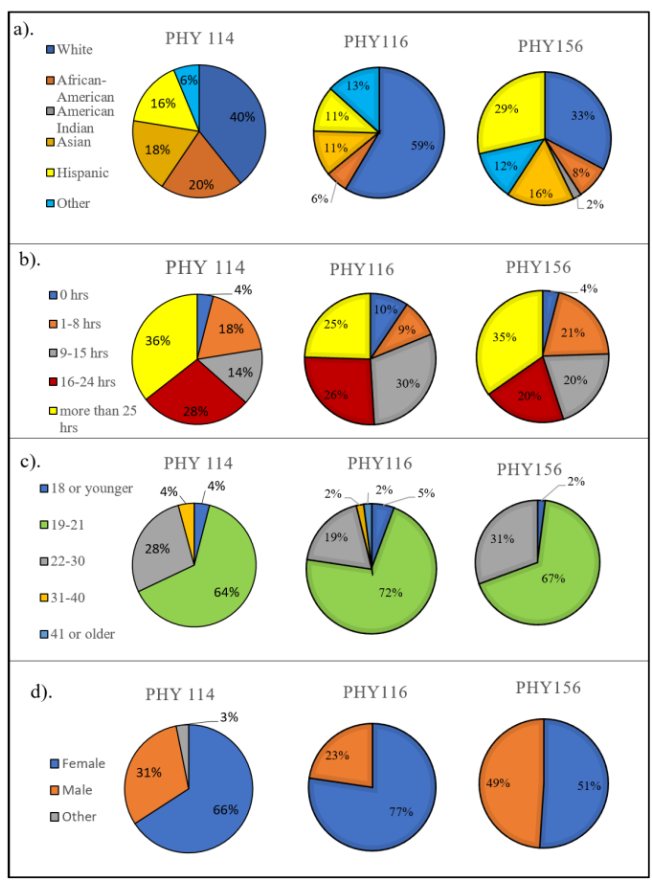

*Fig. 1. Students Demographics: a) shows the class ethnicity, b) shows outside work hours per week, c) shows age groups, and d) shows gender identity.*

Learners' education, participation, commitment, respect, and class engagement in entire online class activities should be addressed. They all can be considered as our main concerns that need to be improved. Especially, when there are students who don't seek help, at least when the instructor thinks they need it, the instructor will feel unhappy about the usage of given opportunities by creating a barrier with the instructor. These classes are personalized by the instructor for students to use instructor materials effectively by following specific class requirements and class ethics. There were a lot of concerns detected from these personalized class structures. Since the classes are online, students are not committed to work hard or spend time effectively, not be honest with their work and use unauthorized or incorrect outside resources that do not match with what instructor taught and provided. They only spend a few minutes and find a quick way to respond to the problems without clear understanding. There are some students even don't engage in homework assignments by learning and effort. Many homework assignments are submitted just to earn allocated points, but there are a lot of students even don't know what questions were given in the homework, because someone else does the work for them. Then, they face quizzes even without knowing that the problems were given earlier. On the other way 20% of students are really focused into given materials and instructor guidelines. Students in this category score almost all assigned points and receive letter grade "A" without any difficulties. By spreading those students to other groups of students can be benefited to others.

Figure 3 shows students respect, fairness, and honesty regarding the assignments guide and class requirements tested in the spring 23 without group work implementation and fall 23 with group work implementation. The quality of the work in addition to students' respect, honesty and fairness can be investigated from this data. There were a lot of issues as can be seen in spring 23. In Spring 2023, homework assignments and class rules are just like group work assignments shown in Figure 2, but it was individualized to each student. Learners' outcomes of the whole class structure and their continues commitments throughout the semester were measured by using grade distribution and percentage of missing assignments as in Figure 4. There were a lot of missing assignments on spring 23. Students were able to discuss homework assignments with peers in spring 23 if they would like to but submit work individually for all assignments on their own effort. The assignments were individualized with all calculations and conclusions by students' college ID integration and breakdown problems to maintain academic honesty as shown in Figure 2 [2].

*C. Expected benefits from group work*

To enhance the student learning outcomes in this personalized class structure and to overcome the above concerns, we decided to cease the homework assignments completely and use one hour class period to conduct groupwork in an effective manner. Creating an effective in class group work concept will benefit students to share their ideas, communicate, and seek help from the instructor as a group. This will also help students to go over class ethics and class requirements together by focusing on instructor expectations and valuing academic honesty and education continuously which will produce the quality work. Group work will benefit students in various ways, a few are listed below.

- Learn from each other in all aspects such as subject matter, social connections, time management, communication, collaboration, organization, sharing, and many more.

- Learn from mistakes and revisit the feedback as a group, learn how to read feedback and track grades through LMS.

- Reduce individual workload, homework and spend class time effectively.

- Revisit class requirements and goals together to cease the penalties and help each other.

- Practice and follow work submissions and due dates effectively and learn time managements habits.

- Chance to compare work and identify what is wrong in their own work. This forces them to seek help.

- Pinpoint where to study and how to study by working with students who really do hard work and follow guidance.

- Get a chance to conclude the observation as a group and open the discussion further.

- Learn to listen and appreciate others' work and ideas.

- Enhance the communication of subject matter, socialize and value their life and education.

- Enhance the encouragement to meet instructors as a group or as an individual to get help.

- Understand the problems that are common among most students, emphasizing that I am not alone.

- Find a connection and relevance to real-life experience in different disciplines.

- Make friendships and connections not only within the college education but also for their future.

Additionally, this will unload the grading time for instructors since it is a group grading. This will be a very effective way to provide feedback as a group and it saves time for such a large class to use for intensive class development and progress.

*D. Structure of the Project and the assignments*

Our expectation from the project is to improve students' engagement, learning outcomes, adhere to guidelines and class ethics, respect, fairness, and honesty of work by introducing an inclusive learning environment with group assignments. We impose culturally responsive pedagogy, diversity inclusion, social justice, experimental learning, and community engagement to learn from each other not only the subject matter but also many other life skills in students' mindsets.

A Google sign-up sheet is provided for students to sign-up for their own choice within the 18 groups of students in the lab section. The number of students per group is limited to 3-4. We conduct all group meetings through Zoom by preassigned break outs rooms according to the sign-up sheet. The instructor helps through Q and A, meeting with groups who need help and switching groups during the session. All submissions are set up in LMS by creating its groups feature that can be accessed by each member of the group to see the work and feedback equally.

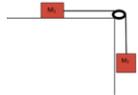

*Fig. 2. An example assignment: Group work assignment with detailed class requirements and directions to follow with sample ID integrated problem.*

A problem set is created for each unit that reflects what students learn in e-lectures and problem-solving sessions by applying the knowledge together to real-life examples. The assignment is accessible to everyone in the class and mitigates all levels of biasing. Assignment requirements are imposed the same way for the other assignments and the same as spring 23 because we would like to measure the respect and commitment towards guidelines by working as a group. The assignment is redesigned to use the last digit of all group members' school IDs instead. Problems are created by integrating those numbers. Assignments will reflect what they learn from the class by opening the discussion to expand students' thinking jointly. Throughout the assignment, students work together to build the path to success in the course without any barriers that can discriminate or create favoritism. The problems are created to mitigate any biasing such as name, gender, age, body shape, ethnicity, origin, diversity, minority, and so on. Relating the problems to real-world applications and asking them to conclude jointly from what they observed and discussed will help them to restate individual ideas. The example assignment is shown in Figure 2.

The list of roles is assigned to members of the group and rotated them for each new assignment to have equal opportunities for all members. If a few members are absent in the sessions, we request them to distribute the roles to the rest of the members as combined roles. Students are supposed to include all names and roles on the final product as shown in Figure 2. All members should agree on the final product and work equally on each question and understand each step of the work. This will help students to contribute equally to all aspects of learning, communicating, organizing, and collaboration. Each student completes an allocated task that contributes to the final group product.

*1) List of roles:*
i. Responsible for distributing, checking, and handling the roles among the group. Ask each member to introduce themselves to the group during the first session. Help to those who need guidance for submitting, scanning, following due dates, tracking grades or anything else related to class work.
ii. Taking attendance, managing time, and arranging extra meetings. Responsible for following and checking the ground rules related to justice, equity, diversity inclusion (JEDI) during the session.
iii. Imposing and explaining the assignment rules and requirements throughout the work and the session. Listen to the instructor very carefully and revisit them with the group as needed.
iv. Writing the product with all other members' agreement and submitting work on time. Managing discussion or asking for any help from the instructor during breakout rooms.

Group work assignments are graded by following the given grading rubric which distributes points for the structure, content, organization, illustrations, mechanism with penalties for late work, not followed requirements and class ethics, and using unauthorizes resources and random formulas. The group submits one final product, and all group members receive the same grade with detailed feedback, regardless of individual contributions.

We listed following JEDI ground rules to provide each student justice, equity, and diversity inclusion. They should be reinforced by the related in charge of the group to make sure every student has an equal opportunity and respect for their work, ideas, and learning.

*2) JEDI ground rules:*
i. One person should speak at a time.
ii. Provide reasons when you make statements.
iii. Raise a hand or signal when you have something to say.
iv. Listen carefully to what other people are saying.
v. Respect other people and their ideas.
vi. Do not interrupt others.
vii. Do not speak too long. Give the opportunity to others.
viii. Should not bias in gender, color, major, body shape, college, and all other diversities and demographics.

## III. RESULTS AND DISCUSSION

We measure the students' outcomes, class progress and quality of the work by implementing effective group work concepts towards physics education. First, we measure respect and commitment toward assignment requirements and class ethics since it was our main concern to receive quality work. By collecting data about how students react to class rules and ethics in all assignments, we were able to measure the progress and compare the measurements to individual homework environment conducted in spring 23. This can be used to understand students' commitment, respect, learning, progress, work quality, and so on. The results are shown in Figure 3. The data shows a huge improvement from the spring 23 due to mandatory groupwork implementation. This impact massively to improve the quality of the work in all classroom assignments and tests.

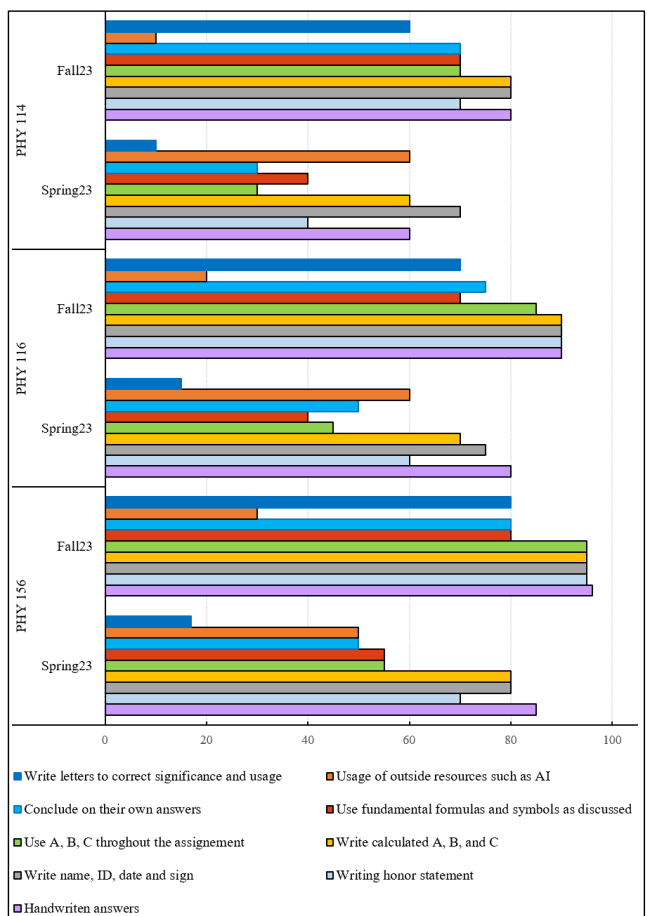

*Fig. 3. Students' respect for all class assignments: Date represents how students react to all class assignments by following specific guidelines to measure student learning and improvement by adapting mandatory group work sessions.*

Further, we tested dishonest usage of ChatGPT [17, 18, 19], AI and unauthorized outside resources for all assignments. This behavior was able to be easily identified by conducting simple investigation on students' work submission and answers' quality check since the class is personalized in a way of using specific symbols, formulas, drawings, statements, and processes [2]. Outside resources usage reduces significantly with implementation of group work in Fall 23 as shown in Figure 2. This shows students commitment and progress of learning materials effectively.

Next, we investigate the grade distributions, and compare the result and improvements of the learning outcomes by working as a group. Detailed work investigations are performed to see the work progress. We investigate how the workload is affected to the grade distribution by unloading homework and loading in-class group work. The results are shown in Figure 4. We investigated the percentage of total final points for the lecture portion of all classes. Since the laboratory part was conducted by different instructors and we did not include that in the comparison. Therefore, the results in Figure 4 will not represent the final letter grades for the students. Although the grade distribution does not show much difference in both semesters, students in the range of 70-92 have been improved in each class. There is significant improvement of submitting work as shown on the right panels in Figure 4. That means working together in some class assignments helps students to manage time and track work and grades effectively.

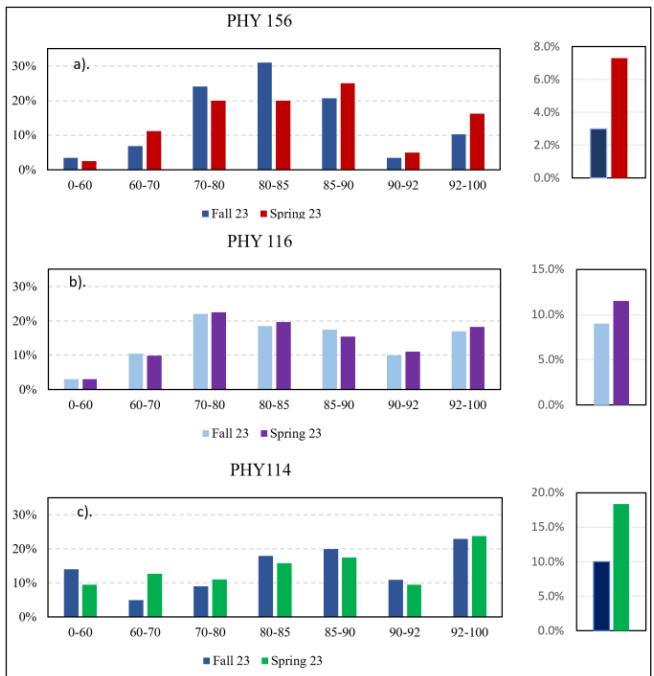

*Fig. 4. Grade distribution: a) shows students grade distribution with percentage of missing assignments on the right for PHY 156. b) and c) represent the same as above for PHY116 and PHY114 respectively. The right panes represent the percentage of all missing assignments during the semester.*

Finally, individual feedback from anonymous surveys is collected and analyzed to evaluate the progress and worthiness of the group work from the student point of view. 75% of the class volunteers for the surveys and around 5% decided to work individually in the middle of the semester due to some discrepancies within groups such as, lack of contribution, carelessness, helpless, unfit, and participation.

Survey results are discussed in Table 1.

*TABLE I. Student Survey results about group work*

| 1. Did you attend almost all group meetings regularly and arrive on time? | |
|---|---|
| Yes, arrive on time | 93.9% |
| Yes, but late always | 3.6% |
| Somewhat and late always | 2.5% |
| Never attend | 0% |
| 2. Do you think you contribute significantly to the success of the project? | |
| Yes | 84.3% |
| No | 1.8% |
| Somewhat | 13.9% |
| 3. Did you feel alone or uncomfortable during group work or see any biasing in your group? | |
| Yes, a lot of biasing happens | 9.4% |
| No, everyone helps each other and no biasing at all | 70.8% |
| Somewhat, because of a few biasing | 19.8% |
| 4. Did your group plays assigned role effectively and rotate to each project? | |
| Yes | 53.5% |
| No | 20.0% |
| Some what | 26.5 |
| 5. Did your group follow the given JEDI ground rules? | |
| Yes, well followed | 80.5% |
| No, not at all | 6.6% |
| Somewhat followed | 11.9% |
| 6. Do you think group work gave you the confidence to do more advanced work in the subject? | |
| Yes, it was very helpful | 66.9% |
| No, it was not helpful | 6.0% |
| No, I prefer to work alone | 27.1% |
| 7. Do you have supportive and encouraging members to learn better? | |
| No, they were unsupportive | 9.2% |
| Yes, they were very supportive | 44.5% |
| Yes, they were somewhat supportive | 44.3 |
| 8. Did group work help you to follow class ethics and requirements? | |
| Yes, I was able to understand them because of working together | 73.2 |
| No, I was able to understand them my self | 26.8 |
| 9. What do you think about breakout rooms, blackboard groups set up by the instructor? | |
| All clear and very effective | 100% |
| Unorganized and hard to follow | 0% |
| 10. Does your group members help you to go back on materials and prepare well | |
| Yes, by working with my group I realize that I need to do more work effectively | 68.7% |
| No, by working with my group I realize that I am the only one working this group | 21.3% |
| No, my group discourage me to do work | 10.0% |
| 11. What is your overall experience by working with a group? | |
| Very impressive and supportive | 40.0% |
| Somewhat okay | 44.5% |
| Unsupportive | 15.5% |

## IV. Conclusion and future plans

Incorporated group work strategies in this paper promote the best practices in all levels of education. Students' enrollment and successful completion of introductory physics courses shows huge improvement.

By participating in group work, students benefited from various life learning skills, communications, organizations, work together, interest on subject matter, value of education, responsibilities, time management, respect each other's and many more. Group work sessions are more likely to be better prepared and show quality work and results when students take individual quizzes.

We share this work because it is easily implemented for any course and impacts students' education and life learning skills. We believe that all students respond favorably to groupwork because they view all peers and instructors as helpers and consider all as free resources to succeed in education.

In conclusion, we found that semester long in-class group work set up with effective strategies to help on each other were helpful for building motivation, increasing learning outcomes, decreasing academic dishonesty, decreasing a number missing work, increasing time management skills, providing equitable and inclusive support, and many more benefits. Since we could not fully achieve our expectations by establishing group work strategies in this first attempt, we are planning to implement new methodologies and strategies to make it more successful in the upcoming year. Among our new strategies, requesting individual work from each member, reimplement roles and rules for the best practice, let be work individually if like to, self or peer gradings, separated work for each meeting, discuss solutions in each meeting to the given set, separate time to value JEDI rules are a few to add.

## Acknowledgment

We acknowledge the College of Staten Island and the Department of Physics and Astronomy for allowing us to remodel the classes and collect all necessary data for this study. We acknowledge the CUNY Office of Faculty Affairs and CUNY Innovative Teaching Academy (CITA) for supporting this research. This research received no financial support.

## Ethical Statement

This research was done according to ethical principles outlined by the Institutional Review Board (IRB) at the College of Staten Island. The project didn't require IRB review because the manuscript was determined to be an educational quality improvement initiative, not research with human subjects. This article does not contain any personal data, and it does not identify any individual. All students have given explicit consent to participate in this research project.